\documentclass[preprint,12pt]{elsarticle}

\usepackage{graphicx}
\usepackage{epsfig}

\usepackage{amssymb}

\journal{Nuclear Physics A}

\begin{document}

\begin{frontmatter}



\title{Deformation effects in the Coulomb breakup of $^{31}$Ne}


\author{Shubhchintak}
\ead{shubhdph@iitr.ernet.in}
\author{R. Chatterjee}
\ead{rcfphfph@iitr.ernet.in}
\address{Department of Physics, Indian Institute of Technology - Roorkee, 247667, INDIA}

\begin{abstract}

We present a fully quantum mechanical theory to study the effects of deformation on various reaction observables in the Coulomb breakup of neutron rich exotic medium mass nuclei on heavy targets within the framework of finite range distorted wave Born approximation by using a deformed Woods-Saxon potential. As an application of this theory, we calculate the one-neutron removal cross section, relative energy spectra, parallel momentum distributions and angular distributions in the breakup of $^{31}$Ne on Pb and Au targets at 234 MeV/u. We suggest ways to put constraints on the large uncertainty in the one-neutron separation energy of $^{31}$Ne and also argue that if $^{31}$Ne
is indeed a halo nucleus then it should be a deformed one.
\end{abstract}


\end{frontmatter}


\section{Introduction}
\label{1}
In the past few decades the study of exotic few body systems has been focused on light fragile systems like halo nuclei. With increasing experimental sophistication, one has now extended this study to the medium mass region where long held concepts in nuclear physics - like that of magic numbers appear to breakdown in some neutron rich nuclei away from the ``valley of stability". In fact, in the region around $N = 20$, strongly deformed nuclei have been found \cite{moto95}. This region, called the ``island of inversion"  \cite{Warburton}, consists of unstable nuclei from $^{30}$Ne to $^{34}$Mg around $N = 20$. The large $B$(E2) values and low lying first excited states suggest a strong deformation in these nuclei \cite{caurier98, utsu99, iwas01, yanagi03}. Furthermore, the inversion between normal-$sd$ and intruder-$pf$ shell has been suggested \cite{doorn09}, which also emphasizes the  deformed structure in these nuclei. It would thus be interesting to study the reactions of these neutron rich nuclei and the effect of their deformed structure on different reaction observables. In this context, the $^{31}$Ne nucleus (with $N = 21$) has attracted great attention as it is speculated to be the heaviest ``halo nucleus", which lies in the ``island of inversion" \cite{Nakamura2}. The theoretical calculations in Refs. \cite{Poves,Descouvemont99} also predict the shell inversion in $^{31}$Ne. Further, there is also a large uncertainty in its one-neutron separation energy ($S_n = 0.29 \pm 1.64$ MeV) \cite {Jurado}, a point which we shall discuss later in this paper.

The one-neutron removal cross section, $\sigma_{n}$, of $^{31}$Ne on $^{12}$C and $^{208}$Pb targets has been measured for the first time by Nakamura {\it et al.} \cite{Nakamura2}, at beam energy of 234 MeV/u. They interpreted the large value of Coulomb breakup cross section by considering the valence neutron in 2$p_{3/2}$ (for $S_n \approx 0.4$ MeV) or 2$s_{1/2}$ (for $S_n \le 0.8$ MeV) and not in 1$f_{7/2}$, as one would have expected from a standard shell model configuration. Recently, Takechi {\it et al.}, \cite{Takechi,Takechi2} measured the interaction cross section, $\sigma_{I}$, of $^{28-32}$Ne on $^{12}$C target at 240 MeV/u. They also explained the large value of $\sigma_{I}$ of $^{31}$Ne by considering $s$- or $p$-wave valence neutron \cite{Takechi2}, which is consistent with Ref. \cite{Nakamura2}.

The ground state spin-parity ($J^{\pi} = 3/2^-$) suggested by these experiments also agree with the theoretical predictions of Refs. \cite{Poves, Descouvemont99}. Furthermore, the theoretical calculations within Glauber and eikonal models \cite{Horiuchi} also suggest the ground state spin-parity of $^{31}$Ne to be ${3/2}^-$ by comparing the calculated value of $\sigma_{n}$ with that of experiment \cite{Nakamura2}. Therefore, one can expect the intruder configuration and deformed structure in $^{31}$Ne. In spite of this, one cannot eliminate the possibility of a $1/2^+$ ground state, as mentioned above, which can cause even more drastic changes to the nuclear structure. To interpret the data \cite{Nakamura2} in terms of deformation, Hamamoto did a Nilsson model calculation using a deformed Woods-Saxon potential \cite{Hamamoto} and suggested that the ground state $1/2^+$ of $^{31}$Ne comes from Nilsson level [200 1/2] for $S_n > 500$ keV with quadrupole deformation parameter $\beta_2 \ge 0.6$, whereas $3/2^-$ ground state comes from [321 3/2] Nilsson level (for $S_n < 200$ keV and 0.40 $\le$ $\beta_2$ $\le$ 0.59) or from [330 1/2] Nilsson level (for $S_n > 200$ keV and 0.22 $\le$ $\beta_2$ $\le$ 0.30). The effect of deformation on the reaction cross section has also been studied in Refs. \cite{Minomo1,Sumi,Minomo2} by using the microscopic optical potential model. They claimed that $^{31}$Ne is strongly deformed ($\beta_2 \approx 0.4$) having a halo structure with ground state spin-parity 3/2$^-$. Recently, a particle-rotor model has been used in Refs.\cite{Urata1,Urata2}, to analyze the experimental data \cite{Nakamura2} and to study the effects of deformation on the reaction cross section. The coupling between the motion of valence neutron and the rotational motion of the deformed core ($^{30}$Ne) was taken into account. The ground state spin-parity suggested was 3/2$^-$ corresponding to core deformation parameter $\beta_2 \approx 0.2$, even though there was a possibility of $J^{\pi}$$=1/2^+$ with $\beta_2 \approx 1$ or $\beta_2 \approx -0.4$. Takechi {\it et al.} \cite{Takechi3}, also analyzed the data \cite{Takechi} by considering the single particle levels in a deformed potential. Using the $\beta_2$ value ($0.6 \pm 0.2$) of the core, deduced from the experimental $B$(E2) data \cite{yanagi03}, it was suggested that $^{31}$Ne has $s$- or $p$-orbital halo structure corresponding to Nilsson level [200 1/2] or [321 3/2], respectively. In fact, all these studies emphasize that the valence neutron in $^{31}$Ne occupies the 2$p_{3/2}$ orbital and not the 1$f_{7/2}$, although the possibility of 2$s_{1/2}$ is not fully excluded. 

In this context, we present a detailed study of the Coulomb breakup of $^{31}$Ne at 234 MeV/u, and suggest a systematic way to incorporate deformation at the transition matrix level within the framework of post form finite range distorted wave Born approximation (FRDWBA). The theory includes the electromagnetic interaction between the fragments and the target to all orders. Furthermore, the breakup contribution from the entire non-resonant continuum (corresponding to all multipoles and relative orbital angular momentum between the fragments) are also accounted for. The uncertainties associated with multipole strength distributions in many other formalisms are also avoided as one needs only the ground state wave function of the projectile as an input. The analytic nature of this theory stems from the fact that pure Coulomb wave functions are used in the calculation and that the dynamics can be analytically evaluated. The theory is then used to study the effects of deformation on various reaction observables using a deformed Woods-Saxon potential. We also identify those observables which are more affected by deformation of the projectile and could be a subject of future experiments.

The paper is organized in the following way. In section 2, we present details of the formalism. Our results are presented in section 3 where we discuss the effects of deformation on the Coulomb breakup cross section, relative energy spectra, parallel momentum distribution (PMD), angular and energy angular distributions, respectively and finally in section 4 we present our conclusions.
\section{Formalism}
\begin{figure}[h]
\centering
\includegraphics[height=6.0cm, clip,width=0.75\textwidth]{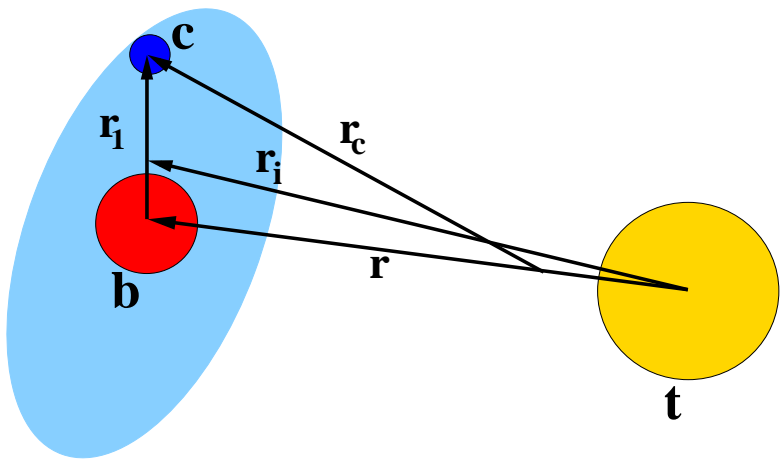} 
\caption{\label{fig1} Three-body Jacobi coordinate system with deformed projectile.}
\end{figure}

We consider the elastic breakup of a two body composite projectile $a$ in the Coulomb field of target $t$ as: $a+t\rightarrow b+c+t$, where projectile $a$ breaks up into fragments $b$ (charged) and $c$ (uncharged). The three body Jacobi coordinate system adopted is shown in Fig.\ref{fig1}. 

The position vectors ${\bf {r}}_{1}$, ${\bf {r}}_{i}$, ${\bf{r}}_{c}$ and ${\bf{r}}$ satisfy the following relations,
\begin{eqnarray}
{\bf{r}}={\bf{r}}_{i}-\alpha{\bf{r}}_{1} ;\hspace{0.2in} {\bf{r}}_{c}=\gamma{\bf{r}}_{1}+\delta{\bf{r}}_{i}\label{a1}.    
\end{eqnarray}
$\alpha$, $\gamma$ and $\delta$ are the mass factors, given by:
\begin{eqnarray}
\alpha =\frac{{m_{c}}}{m_{c}+m_{b}};\hspace{0.2in}  \delta =\frac{m_{t}}{m_{b}+m_{t}};\hspace{0.2in} \gamma=(1-\alpha\delta) \label{a2}
\end{eqnarray}
where, $m_{b}$, $m_{c}$ and $m_{t}$ are the masses of fragments $b$, $c$ and $t$, respectively.

The triple differential cross section for the reaction is related to reduced transition amplitude $\beta_{\ell m}$ as
{\footnote {For more details of the derivation from post form of the DWBA transition amplitude to the reduced transition amplitude one is referred to Ref. \cite{rc_4th}.}}:
\begin{eqnarray}
\frac{d^3\sigma}{dE_{b}d\Omega_{b}d\Omega_{c}} = \frac{2\pi}{\hbar v_{a}}\rho(E_{b},\Omega_{b},\Omega_{c})\sum_{\ell m}|\beta_{\ell m}|^2,\label{a4.1}
\end{eqnarray}
where $v_{a}$ is the $a-t$ relative velocity in the entrance channel and $\rho(E_{b},\Omega_{b},\Omega_{c})$ is the phase space factor \cite{fuchs} appropriate to the three-body final state. 

The reduced transition amplitude, $\beta_{\ell m}$, is given by 
\begin{eqnarray}
\hat{\ell}\beta_{\ell m}({\bf{q}}_{b},{\bf{q}}_{c};{\bf{q}}_{a})= 
\langle \chi_{b}^{(-)*}({\bf{q}}_{b},{\bf{r}})
\chi_{c}^{(-)*}({\bf{q}}_{c},{\bf{r}}_{c})|V_{bc}({\bf{r}}_{1})|\phi_{a}^{\ell m}({\bf r}_1)\chi_{a}^{(+)}({\bf{q}}_{a},{\bf{r}}_{i})\rangle, \label{a4.2}
\end{eqnarray}
where $\hat{\ell} = \sqrt{2\ell+1}$, ${\bf{q}}_{b}$, ${\bf{q}}_{c}$ and ${\bf{q}}_{a}$ are the  wave vectors of $b$, $c$ and ${a}$ corresponding to Jacobi vectors ${\bf r}$, ${{\bf r}_c}$ and ${{\bf r}_1}$, respectively. $\chi^{(-)}_b$ and $\chi^{(-)}_c$ are the distorted waves for relative motions of $b$ and $c$ with respect to $t$ and the center of mass (c.m.) of the $b-t$ system, respectively, with ingoing wave boundary conditions. $\chi_{a}^{(+)}({\bf{q}}_{a},{\bf{r}}_{i})$ is the Coulomb distorted wave of the projectile with outgoing wave boundary conditions. It describes the relative motion of c.m. of projectile with respect to the target. Further, ${\it\phi}_{a}^{\ell m}({\bf r}_1)=u_{\ell}(r_{1})Y_{\ell m}(\bf\hat{r}_{1})$ is the ground state wave function of the projectile with relative orbital angular momentum state $\ell$ and projection $m$ [$u_{\ell}(r_{1})$ is the radial part and $Y_{\ell m}(\bf\hat{r}_{1})$ is the angular part].

$V_{bc}({\bf{r}}_{1})$, in Eq. (\ref{a4.2}), is the interaction between $b$ and $c$ in the initial channel. This is where we introduce deformation. Following Ref. \cite{ham04}, we write the axially symmetric quadrupole-deformed Woods-Saxon potential (without taking the spin-orbit term) as

\begin{eqnarray}
V_{bc}({\bf r}_1) = V_{ws} f(r_1)-\beta_2 k(r_1) Y^{0}_{2}(\hat {\bf r}_1), \label{a4.3}
\end{eqnarray}
where $V_{ws}$ is the depth of spherical Woods-Saxon potential, $\beta_2$ is the quadrupole deformation parameter and
\begin{eqnarray}
f(r_1) = \frac{1}{1+exp(\frac{r_1-R}{a})}, \hspace{0.2in} k(r_1) = RV_{ws} \frac{df(r_1)}{dr_1}. \nonumber
\end{eqnarray}
The first part of the Eq. (\ref{a4.3}) is the spherical Woods-Saxon potential $V_s(r_1) = V_{ws} f(r_1)$ with radius $R = r_0A^{1/3}$. $r_0$ and $a$ being the radius and diffuseness parameters, respectively. Incidentally, this form of the potential, Eq. (\ref{a4.3}), which is essentially a truncated Tayor expansion of the potential $V_{bc}({\bf r}_1)$, has been used earlier in the literature. We however, refer to only one previous reference \cite{faldt}.

Further, the integral form of the reduced amplitude $\beta_{\ell m}$ [Eq. (\ref{a4.2})], in post form FRDWBA, is given by
\begin{eqnarray}
\hat{\ell}\beta_{\ell m}({\bf{q}}_{b},{\bf{q}}_{c};{\bf{q}}_{a})&=&\int \int d{\bf{r}}_{1}d{\bf{r}}_{i}\chi_{b}^{(-)*}({\bf{q}}_{b},{\bf{r}})\chi_{c}^{(-)*}({\bf{q}}_{c},{\bf{r}}_{c})\nonumber\\
&\times& V_{bc}({\bf{r}}_{1})\phi_{a}^{\ell m}({\bf r}_1)\chi_{a}^{(+)*}({\bf{q}}_{a},{\bf{r}}_{i}).\label{a4.5}
\end{eqnarray}

 One replaces $\chi^{(-)}_c$ by a plane wave as there would be no Coulomb interaction between $c$ (uncharged) and $t$. However, despite the deformed potential we still retain, $\phi^{\ell m}_a({\bf r}_1)=u_{\ell}(r_1)Y^{\ell}_m(\hat{{\bf r}_1})$, i.e. the radial wave function $u_{\ell}(r_1)$ is calculated from the undeformed potential [$V_s(r_1)$], so that we can still have analytical results as will be shown subsequently. We emphasize that the deformation parameter ($\beta_2$) has already entered into the theory via $V_{bc}({\bf{r}}_{1})$ in Eq. (\ref{a4.5}). Of course, we also need to acknowledge that apart from the desire to get analytical results, this procedure will be valid if $\beta_2$ is not too large.

We will now try to factorize Eq. (\ref{a4.5}) (a six-dimensional integral) into two integrals - one over ${\bf r}_1$ and the other over ${\bf r}_i$. Our preferred way is to invoke the local momentum approximation (LMA) to factorize $\chi_{b}^{(-)*}({\bf{q}}_{b},{\bf{r}})$ as 
\begin{eqnarray}
\chi_{b}^{(-)}({\bf{q}}_{b},{\bf{r}}) = 
e^{-i\alpha{\bf K}.{\bf r}_1}\chi_{b}^{(-)}({\bf{q}}_{b},{\bf r}_i), \label{a4.6}
\end{eqnarray}
where ${\bf K}$ is the local momentum of the charged core $b$, whose magnitude is
\begin{eqnarray}
K = \sqrt{\frac{2\mu_{bt}}{\hbar^{2}}(E_{bt}-V({\cal R}))}.
\end{eqnarray}
$\mu_{bt}$ is the reduced mass and $E_{bt}$ is the relative energy of the $b - t$ system. $V ({\cal R})$ is the Coulomb potential between core $b$ and the target $t$ at a distance ${\cal R}$. For more details about the LMA, one is referred to Refs. \cite{rc_4th, rs_ng}.

Therefore, in integral form $\beta_{\ell m}$ reduces to
\begin{eqnarray}
\hat{\ell}\beta_{\ell m} &=& \int d{\bf r}_i e^{-i\delta {\bf q}_c.{\bf r}_i}
\chi^{(-)*}_b({\bf q}_b,{\bf r}_i)\chi^{(+)}_a({\bf q}_a,{\bf r}_i)
\nonumber \\
& \times &
\int d{\bf r}_1 e^{-i{\bf Q}.{\bf r}_1} 
V_{bc}({\bf r_1})\phi_{a}^{\ell m}({\bf r}_1) \label{a4.7},
\end{eqnarray}
where, ${\bf Q} = \gamma {\bf q}_c-\alpha{\bf K}$. The first integral in Eq. (\ref{a4.7}), is the dynamics part in the Coulomb breakup and is expressed analytically in terms of the Bremsstrahlung integral \cite{Nordsieck_1954}. The second integral in Eq. (\ref{a4.7}), contains the structure information, and hence the effects of the deformation. At this stage it is worth noting that except for the structure part, the dynamics part remains the same with the case in which we did not consider any deformation \cite{rc_4th}. 

Let us now turn our attention to the structure part:
\begin{eqnarray}
I_f = \int {\bf dr}_1 e^{-i{\bf Q}.{\bf r}_1} V_{bc}({\bf r}_1) u_{\ell}({r}_1) Y^m_{\ell}(\hat {\bf r}_1) \label{a4.8}
\end{eqnarray}
Expanding the exponential appearing in the previous integral, in terms of spherical harmonics:
\begin{eqnarray}
e^{-i{\bf Q.r}_1} = 4\pi\sum_{l_1m_1} i^{-l_1} j_{l_1}(Qr_1)Y^{{m_1}\ast}_{l_1}(\hat {\bf r}_1)Y^{m_1}_{l_1}(\hat {\bf Q}) \label{a4.9}
\end{eqnarray}
Eq. (\ref{a4.8}) can be written as,
\begin{eqnarray}
I_f&=& 4\pi\sum_{l_1m_1} i^{-l_1} Y^{m_1}_{l_1}(\hat {\bf Q}) \int r^2_1 dr_1 j_{l_1}(Qr_1) \nonumber\\
&\times&\int d\Omega_{r_1}V_{bc}({\bf r}_1)Y^{{m_1}\ast}_{l_1}(\hat {\bf r}_1)Y^{m}_{\ell}(\hat {\bf r}_1) u_{\ell}(r_1). \label{a4.10}
\end{eqnarray}
Now, on substituting the value of $V_{bc}({\bf r}_1)$ from Eq. (\ref{a4.3}), the structure part becomes:
\begin{eqnarray}
I_f  &=& 4\pi\sum_{l_1m_1} i^{-l_1} Y^{m_1}_{l_1}(\hat {\bf Q}) \int r^2_1 dr_1 j_{l_1}(Qr_1) \nonumber\\
&\times&\int d\Omega_{r_1} \left[V_s(r_1) -\beta_{2} R V_{ws} Y^{0}_{2}(\hat {\bf r}_1)\frac{df(r_1)}{dr_1}\right]Y^{{m_1}\ast}_{l_1}(\hat {\bf r}_1)Y^{m}_{\ell}(\hat {\bf r}_1) u_{\ell}(r_1) \nonumber\\
   &=& 4\pi\sum_{l_1m_1} i^{-l_1} Y^{m_1}_{l_1}(\hat {\bf Q}) \int r^2_1 dr_1 j_{l_1}(Qr_1)u_{\ell}(r_1)\nonumber\\
&\times&\Bigg[\int  d\Omega_{r_1} V_s(r_1) Y^{{m_1}\ast}_{l_1}(\hat {\bf r}_1)Y^{m}_{\ell}(\hat {\bf r}_1)- \nonumber\\
&& \int d\Omega_{r_1} \beta_{2}R V_{ws}\frac{df(r_1)}{dr_1} Y^{0}_{2}(\hat {\bf r}_1) Y^{{m_1}\ast}_{l_1}(\hat {\bf r}_1)Y^{m}_{\ell}(\hat {\bf r}_1)\Bigg]. ~ \label{a4.11}
\end{eqnarray}
This can be further simplified to,
\begin{eqnarray}
I_f &=& 4\pi\sum_{l_1m_1} i^{-l_1} Y^{m_1}_{l_1}(\hat {\bf Q}) \int r^2_1 dr_1 j_{l_1}(Qr_1)u_{\ell}(r_1) \nonumber\\
&\times&\left[V_s(r_1) \delta_{l_1,\ell}\delta_{m_1,m}-\beta_{2}R V_{ws}\frac{df(r_1)}{dr_1}\int d\Omega_{r_1}Y^{0}_{2}(\hat {\bf r}_1)Y^{{m_1}\ast}_{l_1}(\hat {\bf r}_1)Y^{m}_{\ell}(\hat {\bf r}_1)\right]. \nonumber\\
~\label{a4.12}
\end{eqnarray}
In the limit of $\beta_2 = 0$, the above equation  would simply contain the first term in the square bracket [involving the spherical potential $V_s(r_1)$] with the summation $(l_1,m_1)$ also being taken care of by the delta functions. The resulting expression would be precisely what one would obtain in the case where there is no deformation \cite{rc_4th}.

We now turn our attention to the second term in Eq. (\ref{a4.12}), involving the term $\beta_2$. The integral over three spherical harmonics is given by,

\begin{eqnarray}
I_1=\int d\Omega_{r_1}Y^{0}_{2}(\hat {\bf r}_1)Y^{{m_1}\ast}_{l_1}(\hat {\bf r}_1)Y^{m}_{\ell}(\hat {\bf r}_1)= (-1)^{m_1}\sqrt{\frac{5}{4\pi}}\left[\frac{(2{\ell}+1)(2l_1+1)}{4\pi}\right]^{1/2}\nonumber\\
  \times \left(\begin{array}{ccc}l_1 & 2 & \ell \\ 0 & 0 & 0 \end{array}\right)\left(\begin{array}{ccc}l_1 & 2 & \ell \\ -m_1 & 0 & m \end{array}\right), \label{a4.13}
\end{eqnarray}
with $|\ell-2| < l_1 < |\ell+2|$ and $m_1 = m$. Notice that there would be a limited number of $l_1$ values to be considered, given that $\ell$ is the orbital angular momentum of the projectile ground state.

Substituting Eq. (\ref{a4.13}) in Eq. (\ref{a4.12}), we now obtain
\begin{eqnarray}
I_f &=& 4\pi\sum_{l_1m_1} i^{-l_1} Y^{m_1}_{l_1}(\hat {\bf Q}) \int r^2_1 dr_1 j_{l_1}(Qr_1)u_{\ell}(r_1)\nonumber\\
&\times&\left[V_s \delta_{l_1,\ell}\delta_{m_1,m}-\beta_{2}R V_{ws}\frac{df(r_1)}{dr_1}I_1 \right].\label{a4.14}
\end{eqnarray}
Furthermore, the spherical harmonic $Y^{{m_1}\ast}_{l_1}(\hat {\bf Q})$ appearing in Eq. (\ref{a4.14}), where ${\bf Q} = \gamma {\bf q_c}-\alpha{\bf K}$, can be written in terms of product of two spherical harmonics one depending on $\hat {\bf q}_c$ and the other depending on $\hat {\bf K}$, using  Moshinsky's formula \cite{Moshinsky}:
\begin{eqnarray}
(|{\bf Q}|)^{l_1}Y^{m_1}_{l_1}(\hat {\bf Q}) &=& \sum_{LM_L}\frac{\sqrt{4\pi}}{\hat L}\left(\begin{array}{c} 2l_1+1\\ 2L\end{array}\right)^{1/2}|\alpha K|^{l_1-L}(\gamma q_c)^L \nonumber\\
&\times&\left\langle l_1-L\hspace{0.1in} m_{1}-M_L\hspace{0.1in} L\hspace{0.1in} M_{L}|l_1\hspace{0.1in} m_{1}\right\rangle Y^{m_1-M}_{l_1-L}(\hat {\bf K})Y^{M}_{L}(\hat {\bf q}_c),\nonumber\\
~
\end{eqnarray}
where $\left(\begin{array}{c} 2l_1+1\\ 2L\end{array}\right)$ is the binomial coefficient and ${\hat L } = \sqrt{2L+1}$ with $L$ varying from 0 to $l_1$. Therefore, the structure part Eq. (\ref{a4.14}), can be evaluated and would contain the effect of the deformation of the projectile.

We wish to emphasize once again the analytic nature of our calculation at this point. With the structure part given by Eq. (\ref{a4.14}), the dynamics part in Eq. (\ref{a4.7}) is still given by the Bremsstrahlung integral, which can be analytically solved. 

Therefore, it is interesting to apply this theory to the breakup of a deformed projectile on a heavy target and investigate the effects of deformation on various reaction observables.
\section{Results and discussions}
\subsection{Structure of $^{31}$Ne}
$^{31}$Ne has a low one-neutron separation energy of $S_n = 0.29 \pm 1.64$ MeV \cite{Jurado} or as per the evaluated Nuclear Data Sheets $0.30 \pm 1.6$ MeV \cite{Singh}. The ground state spin-parity of $^{31}$Ne has been quoted
as $3/2^-$ \cite{Poves,Descouvemont99,Minomo2} or $1/2^+$ \cite{Urata1,Takechi3} and therefore we consider two configurations (i) a $2p_{3/2}$ neutron coupled to a $^{30}$Ne$(0^+)$ core and (ii) a $2s_{1/2}$ neutron coupled to a $^{30}$Ne$(0^+)$ core. In both cases, the single particle relative motion wave function for the neutron is constructed by assuming a Woods-Saxon interaction between the valence neutron and the charged core whose depth is adjusted to reproduce the binding energy. With the radius and diffuseness parameters as 1.24 fm and 0.62 fm, respectively, the depths turn out to be 50.41 MeV and 29.88 MeV for two configurations mentioned above (with $S_n = 0.29$ MeV).

\subsection{One-neutron removal cross sections}
\begin{figure}[ht]
\centering
\includegraphics[height=9.5cm, clip,width=0.7\textwidth]{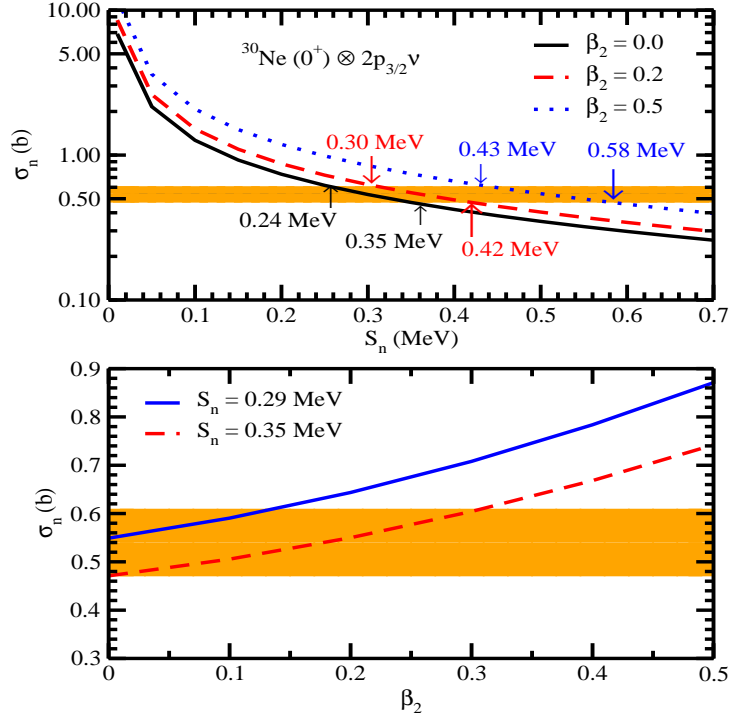}
\caption{\label{fig2} Total cross section for the Coulomb breakup of $^{31}$Ne ($J^{\pi} = 3/2^{-}$) on Pb target at 234 MeV/u beam energy, calculated for different values of $S_n$ (top panel) and for different values of $\beta_2$ (bottom panel). The experimental data shown by the shaded region are from Ref. \cite{Nakamura2}. For more details see text.}
\end{figure}

In Fig. \ref{fig2}, we calculate the one-neutron removal cross section as a function of the one-neutron separation energy (top panel) in the breakup of  $^{31}$Ne on Pb target at 234 MeV/u beam energy, for three cases - (i) no deformation, $\beta_2 = 0.0$ (solid line), (ii) $\beta_2 = 0.2$ (dashed line) and (iii) $\beta_2 = 0.5$ (dotted line). The shaded region corresponds to the experimental data from Ref. \cite{Nakamura2}. One observes that the $p$-wave configuration of the projectile is quite choosy in the range of one-neutron separation energies possible for $^{31}$Ne, with or without deformation. In fact, these calculations, read in conjunction with the Coulomb breakup data \cite{Nakamura2}, can substantially reduce the error bars of the experimental data for the one-neutron separation energy \cite{Jurado,Singh}. To substantiate our point we plot the one-neutron removal cross sections as a function of $\beta_2$ for two values of one-neutron separation energy $S_n = 0.29$ MeV (solid line) and $S_n = 0.35$ MeV (dashed line), in the lower panel of Fig. \ref{fig2}. This calculation helps us narrow down the range of $\beta_2$ possible for $^{31}$Ne. The range of $S_n$ which can reproduce the experimental data corresponding to $\beta_2 = 0.0, 0.2$ and $0.5$ comes out to be 0.24$-$0.35 MeV, 0.30$-$0.42 MeV and 0.43$-$0.58 MeV, respectively. Similarly, from the bottom panel, the value of $\beta_2$ corresponding to $S_n = 0.29$ MeV and 0.35 MeV, comes out to be $\approx$ 0.15 and 0.3, respectively.
\begin{figure}[ht]
\centering
\includegraphics[height=9.5cm, clip,width=0.7\textwidth]{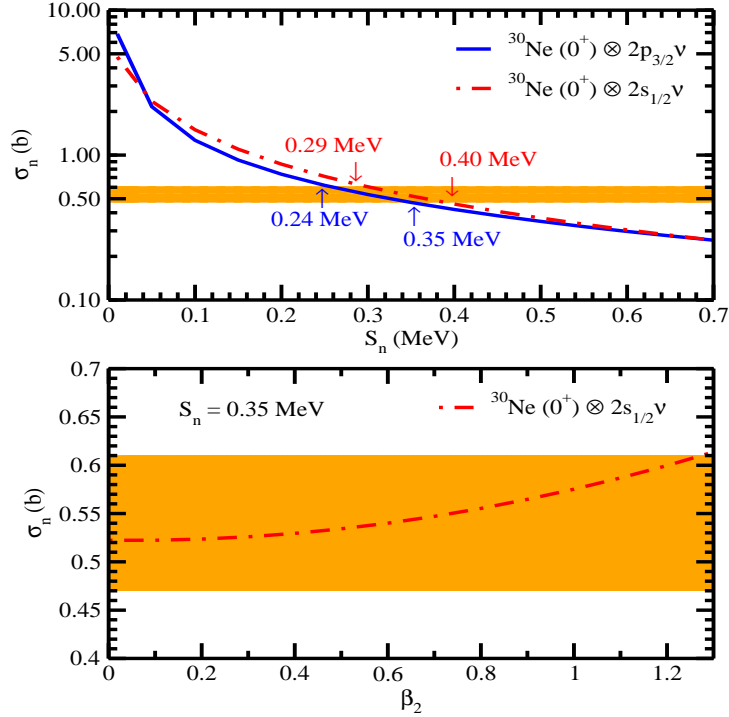}
\caption{\label{fig3} Total cross section for the Coulomb breakup of $^{31}$Ne (without deformation) for $J^{\pi} = 1/2^{+}$ (dot-dashed line) on Pb target at 234 MeV/u beam energy, calculated for different values of $S_n$ (top panel) which is compared with the cross section (without deformation) for $J^{\pi} = 3/2^{-}$ (solid line). The bottom panel shows the cross section for the $J^{\pi} = 1/2^{+}$ configuration of $^{31}$Ne (dot-dashed line) calculated for different values of $\beta_2$. The experimental data shown by the shaded region are from Ref. \cite{Nakamura2}. For more details see text.}
\end{figure}

However, if indeed the spin-parity of $^{31}$Ne is $1/2^+$ (with $^{30}$Ne$(0^+)\otimes2s_{1/2}\nu$ configuration), as some authors suggest, instead of $3/2^+$, we need to check if the available data supports it. In Fig. \ref{fig3}, we plot the one-neutron removal cross section in the breakup of $^{31}$Ne on Pb target at 234 MeV/u beam energy with a \textit{s}-wave configuration, without deformation (dot-dashed line), for $^{31}$Ne as a function of the one-neutron separation energy (top panel) and also as a function of $\beta_2$ (bottom panel) for $S_n = 0.35$ MeV. The solid line in the top panel of Fig. \ref{fig3}, is the calculation with \textit{p}-wave of $^{31}$Ne, which we have included for the purpose of comparison. Indeed, the present one-neutron removal cross section data \cite{Nakamura2} (shaded region in Fig. \ref{fig3}) of $^{31}$Ne does not conclusively rule out the $1/2^+$ spin-parity of the ground state of $^{31}$Ne. In fact, this would also be the motivation to calculate and also measure more exclusive observables in the breakup of $^{31}$Ne. However, this unnatural spin-parity state can have a large $\beta_2$ ($\sim$ 1.2) as it is clear from the bottom panel of Fig. \ref{fig3}, which is consistent with the value of $\beta_2$ ($\sim$ 1) suggested by Urata {\it et al.} \cite{Urata1}.
\subsection{Relative energy spectra}

\begin{figure}[ht]
\centering
\includegraphics[height=9.5cm, clip,width=0.8\textwidth]{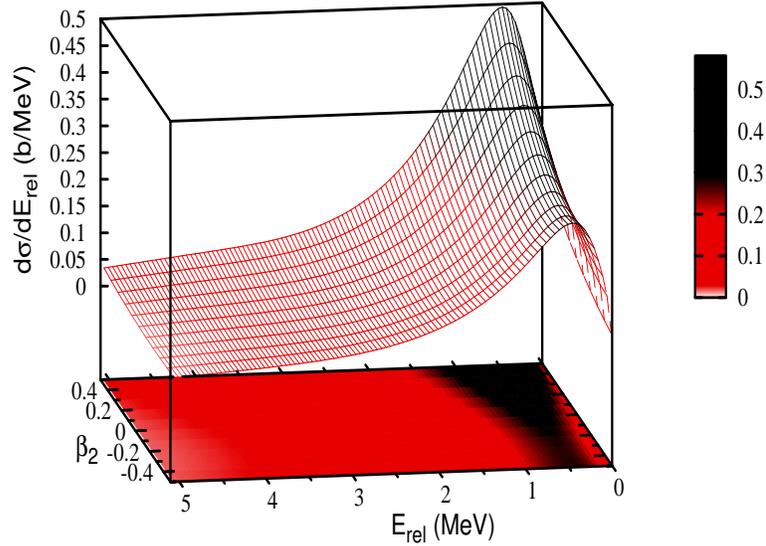} \hskip 24pt
\caption{\label{fig4} Relative energy spectra for the Coulomb breakup of $^{31}$Ne on Pb target at 234 MeV/u beam energy corresponding to different values of $\beta_2$ at $S_n = 0.29$ MeV.}
\end{figure}

In Fig. \ref{fig4}, we show the relative energy spectra in the pure Coulomb breakup of $^{31}$Ne on a Pb target at 234 MeV/u beam energy and simultaneously as a function of $\beta_2$, for the $^{30}$Ne$(0^+)\otimes 2p_{3/2}\nu$ configuration of $^{31}$Ne with $S_n = 0.29$ MeV. The peak position is sensitive to the projectile configuration and the peak height depends on the quadrupole deformation
parameter. It is well known that the peak position of the d$B$(E1)/dE spectra (and also the relative energy spectra) is dependent on the binding energy of the projectile. In fact, the analytic dependence of the peak position as a function of binding energy for a projectile ground state orbital angular momentum $\ell$ is shown in Refs. \cite{nagu,typel_ba}.

Now given the fact that in $^{31}$Ne one has a large uncertainty in its one-neutron separation energy, the measurement of its relative energy spectra and thereby its peak position will go a long way in reducing this uncertainty \cite{rc_fr}. Once this is clarified, calculations of the type shown in Fig. \ref{fig4} can also be used to deduce the quadrupole deformation parameter, which also affects the height of the peak in the relative energy spectra. 

\subsection{Parallel momentum distribution}

\begin{figure}[ht]
\centering
\includegraphics[height=9.5cm, clip,width=0.75\textwidth]{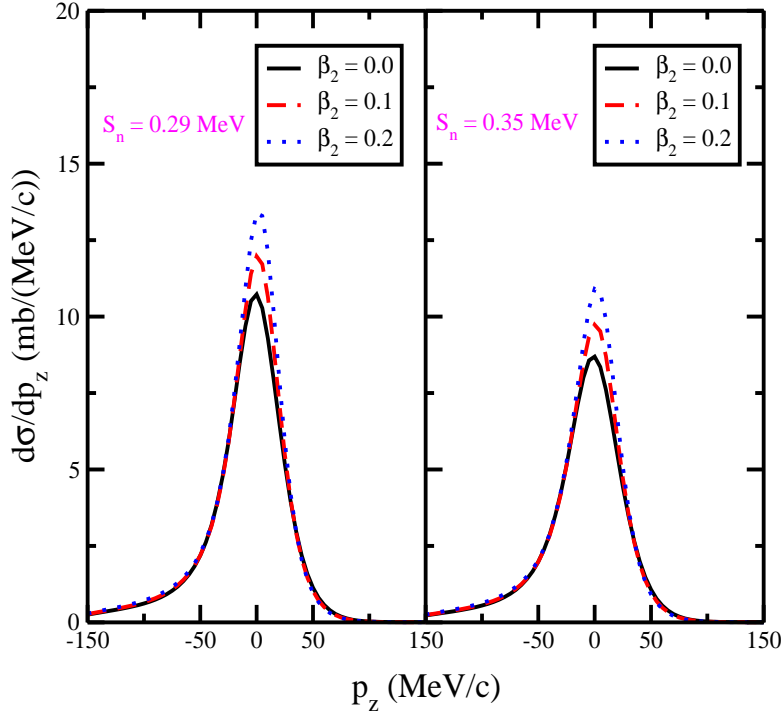}
\caption{\label{fig5} Parallel momentum distribution of $^{30}$Ne calculated from the Coulomb breakup of $^{31}$Ne on Au target at 234 MeV/u beam energy for (a) $S_n = 0.29$ MeV (left panel) and (b) $S_n = 0.35$ MeV (right panel). For more details see text.}
\end{figure}
\begin{table}[ht]
\begin{center}
\caption{\label{tab:table4.1} Effect of deformation on the FWHM of PMD of $^{30}$Ne, obtained in the Coulomb breakup of $^{31}$Ne on Au target at 234 MeV/u, corresponding to two different values of $S_n$.}
\begin{tabular}{|c|c|c|}
\hline\hline
$S_n$ (MeV) & $\beta_2$ &  FWHM (MeV/c)  \\
\hline 
     & 0.0 & 51.24 \\
     & 0.1 & 47.67  \\
0.29 & 0.2 & 45.09  \\
     & 0.3 & 43.62 \\
     & 0.4 & 42.15 \\
     & 0.5 & 42.15 \\
\hline
     & 0.0 & 55.38 \\
     & 0.1 & 50.48  \\
0.35 & 0.2 & 48.03  \\
     & 0.3 & 46.07  \\
     & 0.4 & 44.60 \\
     & 0.5 & 44.60  \\     
\hline
\hline
\end{tabular}
\end{center}
\end{table}

In Fig. \ref{fig5}, we show the parallel momentum distribution of the charged $^{30}$Ne fragment in the  Coulomb breakup of $^{31}$Ne on Au target at 234 MeV/u beam for values of $S_n = 0.29$ MeV (left panel) and $S_n = 0.35$ MeV (right panel). In both cases, the solid, dashed and dotted lines correspond to $\beta_2 = 0.0$, $\beta_2 = 0.1$ and $\beta_2 = 0.2$, respectively. It is interesting to note that the effects of deformation would be pronounced near the peaks, which correspond to the beam velocity momenta. This is encouraging as it seems to suggest that Coulomb breakup experiments would be quite suited to investigate the deformation effects in $^{31}$Ne.

In Table \ref{tab:table4.1}, we show the FWHM of the PMD for a range of $\beta_2$ values with $S_n$ ( = 0.29 MeV and 0.35 MeV) as a parameter. It is hardly surprising that at these low binding energies the FWHM is low even for the case when we have no deformation. What is interesting, however, to note is that the FWHM seems to saturate for $\beta_2 \approx 0.4$ and is very near to values reported for ``established" halo nuclei like $^{11}$Be. Our calculations thus suggest that if $^{31}$Ne is indeed a halo nucleus, it would essentially be a deformed one.

\subsection{Angular distributions}

\begin{figure}[ht]
\centering
\includegraphics[height=9.5cm, clip,width=0.75\textwidth]{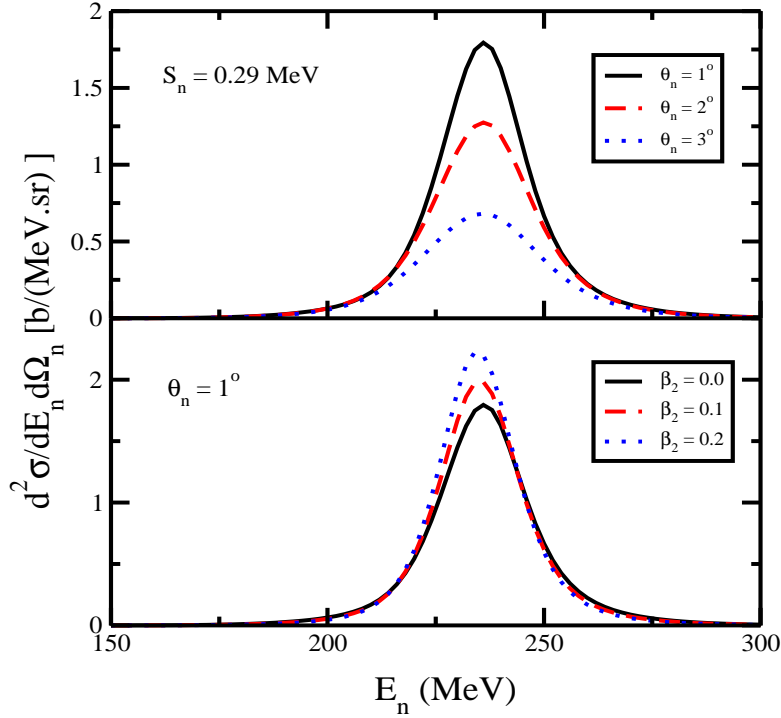} 
\caption{\label{fig6} Neutron energy-angular distribution for the Coulomb breakup of $^{31}$Ne on Au target at 234 MeV/u beam energy calculated at (a) three different angles ($\theta_n$) i.e. at $1^o$, $2^o$ and $3^o$, (b) at $\theta_n = 1^o$, but with different values of $\beta_2$. For more details see text.}
\end{figure}
\begin{figure}[ht]
\centering
\includegraphics[height=9.0cm, clip,width=0.75\textwidth]{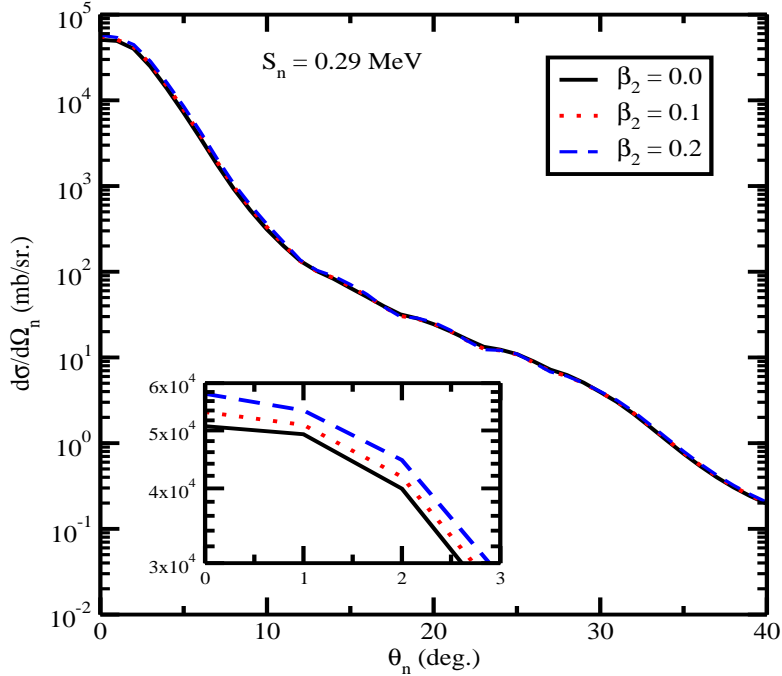}
\caption{\label{fig7} Angular distribution for the Coulomb breakup of $^{31}$Ne on Au target at 234 MeV/u beam energy calculated  at $S_n = 0.29$ MeV. The inset in the figure is the same plot at very forward angles. For more details see text.}
\end{figure}

In Fig. \ref{fig6} (top panel), we show the neutron energy-angular distribution for the Coulomb breakup of $^{31}$Ne on Au target at 234 MeV/u beam energy calculated at three forward angles, $\theta_n = 1^o$ (solid line), $2^o$ (dashed line) and $3^o$ (dotted line), without any deformation of the projectile. The bottom panel of Fig. \ref{fig6}, is a calculation at a forward angle,
$\theta_n = 1^o$ for different values of the deformation parameter $\beta_2$. The solid, dashed and dotted lines correspond to $\beta_2 = 0.0$, $\beta_2 = 0.1$ and $\beta_2 = 0.2$, respectively. It is evident that the effects of deformation will be more near the peaks of the distributions.
It is interesting to observe that the peak position in Fig. \ref{fig6} is quite close to the beam velocity energy, both for the case with and without deformation. This would suggest that there would be no post-acceleration of the charged fragment, a feature which is quite similar to the one with $^{11}$Be \cite{rc_4th}.

In Fig. \ref{fig7}, we present the neutron angular distribution in the breakup
of $^{31}$Ne on Au target at 234 MeV/u beam energy for three values
of the deformation parameter $\beta_2 = 0.0$ (solid line), 0.1 (dotted line)
and 0.2 (dashed line). The inset in the figure is the same plot at very
forward angles. The grazing angle for this reaction is $1.2^o$. The results seem to indicate that the effect of deformation
would be visible essentially at small scattering angles.

We also wish to point out that along with the relative energy spectra one can also put limits on the one-neutron separation energy of these nuclei from angular momentum measurements. It is supposed to be less model dependent \cite{Nak, Nakamura_kondo} than the method dependent on the peak position of the $B$(E1) spectrum, which in some cases could have higher order excitations. We however, wish to emphasize that our relative energy spectrum contains the contribution from the entire non-resonant continuum.

\section{Conclusions}
In this paper, we have extended the quantal theory of Coulomb breakup within the ambit of the FRDWBA to include deformations in projectiles in a simple manner. The formalism retains the analytical flavor of the calculation with the transition amplitude being factorized into two parts - the dynamics and the structure part. The structure part contains the deformation parameter and the dynamics part of the problem can be expressed in terms of the Bremsstrahlung integral - which can be analytically evaluated. This has therefore opened a route to investigate the breakup of deformed neutron rich projectiles in the Coulomb field of a heavy target.

We have used the theory to investigate the breakup of $^{31}$Ne on Pb and Au at 234 MeV/u. We have calculated the one-neutron removal cross section and have compared it with the available data which seems to favor a $3/2^-$ ground state spin-parity for $^{31}$Ne. However, the present data cannot completely rule out the possibility of a $1/2^+$ ground state although that require a very large quadrupole deformation for $^{31}$Ne. This was the motivation in calculating more exclusive reaction observables in the Coulomb breakup of $^{31}$Ne. The peak position of the relative energy spectra was seen to depend on the projectile configuration and deformation parameter. This trend was also confirmed in the parallel momentum distribution of the charged fragment in the breakup of $^{31}$Ne. The FWHM was seen to approach those of halo nuclei for quadrupole deformation, $\beta_2 \approx 0.4$, which confirms the fact that if $^{31}$Ne  is a halo nucleus it is also a deformed one. The effect of deformation was also studied in the angular distributions of the neutron in the final channel and was seen to be significant at very forward angles.

In all these reaction observables we have thus identified regions where the effects of deformation of the projectile would play a vital role. Our calculations, therefore, could also serve as a motivation for future experiments with exotic nuclei in the deformed medium-mass region of the nuclear chart.

In retrospect, one can argue that for a more universal application of our theory, relativistic effects need to be taken into consideration, especially at very high beam energies.  However, a fully relativistic breakup reaction
theory, including proper relativistic dynamics, is still a long way off. Initial efforts in this direction involving
a modified Klein-Gordon equation has already been reported in Ref. \cite{bert_rel},
where the relativistic corrections to the breakup of $^{8}$B 
on Pb were studied
at various beam energies. In fact, for beam energies from $100-250$
MeV/u, these corrections were shown to be in the range of $10-15\%$. Similar
effects were also predicted in Ref. \cite{ogata_rel} where breakup of $^{8}$B
and $^{11}$Be on Pb were studied at $100-250$ MeV/u, using a continuum-discretized
coupled-channels method with eikonal approximation. Interestingly, they also
found that the effects of relativistic corrections to the nuclear potentials,
in their study, were not significant. 

Qualitatively therefore, a similar range as predicted in Ref. \cite{bert_rel, ogata_rel} can be attributed to relativistic effects, in the case of $^{31}$Ne breaking up on heavy targets. However, unless large systematic and statistical uncertainties are reduced in many experiments
it may not be possible to positively discern these effects. Nevertheless,
efforts towards a fully relativistic theory will be a very welcome step in
confronting data due to emanate from new generation of radioactive ion beam
facilities being operated in various parts of the world.

~\\
\noindent{\bf Acknowledgements}

This text presents results from research supported by the Department of Science and Technology, Govt. of India, (SR/S2/HEP-040/2012). [S] also thanks the MHRD, Govt. of  India, for a research fellowship.


\end{document}